\documentclass[final,numbers,times,12pt]{elsarticle}

\usepackage{amsfonts,amstext,amsmath,amssymb}

\usepackage{paralist}
\usepackage{setspace}

\usepackage{hyperxmp}
\usepackage[%
    bookmarks=false,         
    unicode=false,          
    pdftoolbar=true,        
    pdfmenubar=true,        
    pdffitwindow=false,     
    pdfstartview={FitH},    
    pdftitle={A Note on Efficient Performance Evaluation of the Cumulative Sum Chart and the Sequential Probability Ratio Test},    
    pdfauthor={Aleksey S. Polunchenko},     
    pdfsubject={A Note on Efficient Performance Evaluation of the Cumulative Sum Chart and the Sequential Probability Ratio Test},   
    pdfcreator={Aleksey S. Polunchenko},   
    pdfkeywords={Control charts,Cumulative Sum chart,Integral equations,Sequential analysis,Sequential Probability Ratio Test,Quality control}, 
    pdfnewwindow=true,      
    colorlinks=false,       
    linkcolor=red,          
    citecolor=green,        
    filecolor=magenta,      
    urlcolor=cyan           
]{hyperref}

\usepackage[utf8]{inputenc}
\usepackage[T1]{fontenc}
\usepackage{textcomp}

\biboptions{sort&compress}

\newcommand{\ignore}[1]{}

\renewcommand{\Pr}{\mathbb{P}} 
\DeclareMathOperator{\EV}{\mathbb{E}} 

\DeclareMathOperator{\LR}{\Lambda}

\DeclareMathOperator{\ARL}{ARL}
\DeclareMathOperator{\ASN}{ASN}
\DeclareMathOperator{\OC}{OC}

\renewcommand{\le}{\leqslant} 
\renewcommand{\ge}{\geqslant}

\newtheorem{theorem}{Theorem}[section]
\newtheorem{lemma}[theorem]{Lemma}

\usepackage[norefs,nocites]{refcheck}
\usepackage{silence}
\journal{Applied Stochastic Models in Business and Industry}

\begin{document}

\begin{frontmatter}



\title{\large{\bf\uppercase{A Note on Efficient Performance Evaluation of the Cumulative Sum Chart and the Sequential Probability Ratio Test}}}


\author{Aleksey\ S.\ Polunchenko\corref{cor-author}}
\ead{aleksey@binghamton.edu}
\ead[url]{http://www.math.binghamton.edu/aleksey}
\cortext[cor-author]{Address correspondence to A.S.\ Polunchenko, Department of Mathematical Sciences, State University of New York (SUNY) at Binghamton, 4400 Vestal Parkway East, P.O. Box 6000, Binghamton, NY 13902--6000, USA; Tel: +1 (607) 777-6906; Fax: +1 (607) 777-2450; Email:~\href{mailto:aleksey@binghamton.edu}{aleksey@binghamton.edu}}
\address{Department of Mathematical Sciences, State University of New York (SUNY) at Binghamton\\Binghamtom, NY 13902--6000, USA}

\begin{abstract}
We establish a simple connection between certain {\em in-control} characteristics of the CUSUM Run Length and their {\em out-of-control} counterparts. The connection is in the form of paired integral (renewal) equations. The derivation exploits Wald's likelihood ratio identity and the well-known fact that the CUSUM chart is equivalent to repetitive application of Wald's SPRT. The characteristics considered include the entire Run Length distribution and all of the corresponding  moments, starting from the zero-state ARL. A particular {\em practical} benefit of our result is that it enables the in- and out-of-control characteristics of the CUSUM Run Length to be computed {\em concurrently}. Moreover, due to the equivalence of the CUSUM chart to a sequence of SPRTs, the ASN and OC functions of an SPRT under the null and under the alternative can {\em all} be computed {\em simultaneously} as well. This would double up the efficiency of any numerical method one may choose to devise to carry out the actual computations. 
\end{abstract}


\end{frontmatter}

\section{Introduction}
\label{sec:intro}

It cannot be disputed that Wald's~\cite{Wald:Book47} likelihood ratio identity is one of the fundamental methodological tools in all of {\em theoretical} sequential analysis. The powerful change-of-probability-measure technique essentially enabled the proof of nearly every classical result in the areas of sequential hypotheses testing and sequential (quickest) change-point detection: strong optimality of Wald's~\cite{Wald:Book47} Sequential Probability Ratio Test (SPRT) first proved by Wald and Wolfowitz~\cite{Wald+Wolfowitz:AMS1948} (see also~\cite{Wolfowitz:AMS1966}) and then also re-established, e.g., by Matthes~\cite{Matthes:AMS1963} and, notably, by Le Cam, whose proof may be found in~\cite{Lehmann:Book1959}; exact minimaxity (in the sense of Lorden~\cite{Lorden:AMS71}) of Page's~\cite{Page:B54} Cumulative Sum (CUSUM) ``inspection scheme'' established by Moustakides~\cite{Moustakides:AS86} (although an alternative, viz. game-theoretic, proof was also later offered by Ritov~\cite{Ritov:AS90}); exact Bayesian optimality of Shiryaev's~\cite{Shiryaev:SMD61,Shiryaev:TPA63} detection procedure shown in~\cite{Shiryaev:SMD61,Shiryaev:TPA63,Shiryaev:Book78}; exact maximum-probability-type optimality of the Shewhart's~\cite{Shewhart:JASA1925,Shewhart:Book1931} $\bar{X}$-chart proved in~\cite{Pollak+Krieger:SA2013} and in~\cite{Moustakides:SA2014}; and exact multi-cyclic optimality of the Shiryaev--Roberts procedure---to name a few; the Shiryaev--Roberts detection procedure emerged from the independent work of Shiryaev~\cite{Shiryaev:SMD61,Shiryaev:TPA63} and Roberts~\cite{Roberts:T59}---hence, the name,---and its multi-cyclic optimality was established in~\cite{Pollak+Tartakovsky:ISITA2008,Pollak+Tartakovsky:SS09} and in~\cite{Shiryaev+Zryumov:Khabanov2010}. For a recent survey of the state-of-the-art in {\em theoretical} sequential analysis, see, e.g.,~\cite{Polunchenko+Tartakovsky:MCAP2012} or~\cite{Tartakovsky+etal:Book2014}, and the references therein.

More recently, however, in~\cite{Polunchenko+etal:SA2014,Polunchenko+etal:ASMBI2014} the technique was put to a different, more {\em applicative} use: to improve the accuracy and efficiency of the numerical method the authors of these papers developed to compute the performance of the so-called Generalized Shiryaev--Roberts detection procedure; the Generalized Shiryaev--Roberts procedure was proposed in~\cite{Moustakides+etal:SS11} as a headstarted version of the classical Shiryaev--Roberts procedure, and the motivation to headstart the latter was drawn from the seminal work of Lucas and Crosier~\cite{Lucas+Crosier:T1982} where it was proposed to headstart the CUSUM chart. The aim of this work is to extend the ideas laid out in~\cite{Polunchenko+etal:SA2014,Polunchenko+etal:ASMBI2014} beyond the Generalized Shiryaev--Roberts procedure, viz. to the CUSUM chart and the SPRT; the possibility of such an extension was previously entertained in~\cite[Section~5]{Polunchenko+etal:SA2014}. Specifically, in this work we employ Wald's~\cite{Wald:Book47} likelihood ratio identity and establish a connection between a host of {\em in-control} characteristics of the CUSUM Run Length and their {\em out-of-control} counterparts. The connection is in the form of coupled integral (renewal) equations, and the derivation utilizes the well-known observation first made by Page~\cite{Page:B54} that the CUSUM chart is equivalent to repetitive application of the SPRT (with properly selected initial score and control bounds). The Run Length characteristics considered include the entire distribution and all of the corresponding moments, starting from the standard zero-state Average Run Length (ARL). On the {\em practical} side, the obtained connection enables {\em concurrent} evaluation of the in- and out-of-control characteristics of the CUSUM Run Length. This would double up the efficiency of any numerical method one may devise to compute the performance of the CUSUM chart (through solving the corresponding integral equations). The efficiency improvement would be of an even greater magnitude for the {\em two-sided} CUSUM chart, also proposed by Page~\cite[Section~3]{Page:B54}. Moreover, thanks to the observation first made by Page~\cite{Page:B54} that the CUSUM chart is equivalent to a sequence of SPRTs, the Average Sample Number (ASN) and the Operating Characteristic (OC) functions of an SPRT under the null and under the alternative can {\em all} be computed {\em simultaneously} as well, again with the aid of the main result obtained in the sequel. Hence, in a sense, this work is an attempt to bridge the gap between the theory and applications of sequential analysis.

It is worth recalling that the need to evaluate the performance of the CUSUM chart (or that of the SPRT, or any other control chart for that matter) {\em numerically} is dictated by the fact that the corresponding characteristics (e.g., the zero-state ARL, the ASN function, or the OC function) are governed by integral (renewal) equations that seldom allow for an analytical solution; cases where an analytic closed-form solution {\em is} possible are offered, e.g., in~\cite{Regula:PhDThesis1975,Gan:SS1992,Vardeman+Ray:T1985,Knoth:PhDThesis1995,DeLucia+Poor:IEEE-IT1997,Knoth:SA1998,Mazalov+Zhuravlev:PCS2002} for the CUSUM chart, in~\cite{Dvoretzky+etal:AMS1953,Albert:AMS1956,Kiefer+Wolfowitz:NRLQ1956,Schorr:U1967,Kohlruss:SA1994} for the SPRT, in~\cite{Novikov:TPA1990,Gan:JQT1998,Polunchenko+etal:ShLnkJAS2013} for the Exponentially Weighted Moving Average (EWMA) chart (introduced by Roberts~\cite{Roberts:T59}), and in~\cite{Pollak:AS85,Kenett+Pollak:IEEE-TR1986,Mevorach+Pollak:AJMMS91,Polunchenko+Tartakovsky:AS10,Tartakovsky+Polunchenko:IWAP10,Polunchenko+Tartakovsky:MCAP2012,Du+etal:SMTA2015} and~\cite[Chapter~4]{Du:PhD-Thesis2015} for the Generalized Shiryaev--Roberts procedure. Since control charts' performance evaluation is a persistent problem in applied sequential analysis (notably in quality control), numerical treatment of the corresponding integral equations has {\it de~facto} become a separate research field, and the literature on the subject is vast indeed. For a recent survey of the state-of-the-art in the field, see, e.g.,~\cite{Li+etal:JSCS2014}. By and large, two types of approaches can be distinguished: randomized (i.e., simulation) and deterministic. For specific examples, see, e.g.,~\cite{Page+Cox:MPCPhS1954,Brook+Evans:B1972,Champ+Rigdon:CommStat1991},~\cite{Moustakides+etal:SS11},~\cite{Polunchenko+etal:SA2014,Polunchenko+etal:ASMBI2014} and~\cite[Chapter~3]{Du:PhD-Thesis2015}. To get a clear picture as to the capabilities of a control chart in a concrete observations model, the chart's performance needs to be evaluated both in the in-control regime as well as in the out-of-control regime. The problem, however, is that the in- and out-of-control regime equations are usually treated {\em separately}, which is obviously inefficient. The reason, in part, is that (apparently) there is no a simple and explicit relationship between the in- and out-of-control regime equations. This work proves otherwise, if the chart of interest is either the CUSUM chart or the SPRT. Specifically, using the result obtained in the sequel, the {\em in-} and {\em out-of-control} characteristics of the CUSUM chart and those of the SPRT under the null and under the alternative can {\em all} be computed {\em simultaneously} and {\em irrespective} of which particular numerical method---whether randomized or deterministic---is used to carry out the actual calculations.

The rest of the paper is organized as follows. Section~\ref{sec:preliminaries} provides the necessary preliminary background. The centerpiece of the paper is Section~\ref{sec:main-results} which is where we first establish our main result and then also explain how exactly it can be used to compute the zero-state in- and out-of-control ARLs of the CUSUM chart and the SPRT's ASNs and OCs under the null and under the alternative---all in one run of whatever numerical method one may devise to perform the computations. Section~\ref{sec:conclusion} draws a line under the entire paper.

\section{Preliminaries}
\label{sec:preliminaries}

To fix ideas, suppose we wish to ``sense'' whether or not the common probability distribution function (pdf) of a ``live''-sampled series of independent observations $X_1,X_2,\ldots$ has changed from $f_0(x)$ initially to $f_1(x)\not\equiv f_0(x)$; in the quality control literature, the densities $f_0(x)$ and $f_1(x)$ are customarily referred to as the on- and off-target distributions, respectively. Since its inception, Page's~\cite{Page:B54} CUSUM ``inspection scheme'' has been just {\em the} tool for the job. The CUSUM scheme flags a alarm at sample number $\mathcal{C}_{h}\triangleq\min\{n\ge1\colon W_n\ge h\}$, where $h>0$ is a control limit (which is selected so as to achieve a desired level of the ``false positive'' risk), $\{W_n\}_{n\ge0}$ is the CUSUM statistic defined as $W_n\triangleq\max\{0,W_{n-1}+\log\LR_n\}$, $n\ge1$, with $W_0=0$, and $\LR_n\triangleq f_1(X_n)/f_0(X_n)$ is the instantaneous likelihood ratio (LR) for the $n$-th data point $X_n$. The (random) stopping time $\mathcal{C}_{h}$ is often referred to as the Run Length, for it literally is the length of a single run of the CUSUM statistic $\{W_n\}_{n\ge0}$ before it exits the strip $[0,h)$ through the control limit $h>0$. For simplicity, we shall assume here and throughout the paper that $\LR_1$ is absolutely continuous, although at an additional effort the case of purely nonarithmetic $\LR_1$ can be handled as well.

A more general version of the CUSUM chart, viz. one proposed by Lucas and Crosier~\cite{Lucas+Crosier:T1982}, assumes that the CUSUM statistic $\{W_n\}_{n\ge0}$ is started not off zero, but off a deterministic point $W_0=w\in[0,h)$. This point is a parameter referred to as either the headstart or the ``initial score''. Formally, the respective {\em Generalized} CUSUM Run Length is defined as
\begin{align}\label{eq:GenCS-T-def}
\mathcal{C}_{h}^w
&\triangleq
\min\{n\ge1\colon W_n^w\ge h\},\; h>0,
\end{align}
where the generalized CUSUM statistic $\{W_n^w\}_{n\ge0}$ admits the recurrence
\begin{align}\label{eq:GenCS-W-def}
W_n^w
&\triangleq
\max\{0,W_{n-1}^w+\log\LR_n\},\; n\ge1,\, W_0^w=w\in[0,h),
\end{align}
and it is apparent that setting $w=0$ reduces the Generalized CUSUM chart to the classical one introduced by Page~\cite{Page:B54}. From now on we shall concentrate exclusively on the Generalized CUSUM chart~\eqref{eq:GenCS-T-def}-\eqref{eq:GenCS-W-def}, although, for brevity and without loss of generality, we shall refer to it as simply the CUSUM chart. We note that, as a parameter of the chart, the headstart $w\in[0,h)$ directly affects the Run Length's characteristics, just as does the control limit $h>0$. The effect of the headstart on the performance of the CUSUM chart~\eqref{eq:GenCS-T-def}-\eqref{eq:GenCS-W-def} was thoroughly studied in~\cite{Lucas+Crosier:T1982}.

The two most popular metrics used to quantitatively assess the performance of the CUSUM chart are the zero-state in- and out-of-control ARLs, conventionally denoted as $\ARL_0(w;h)$ and $\ARL_1(w;h)$, respectively; the $0$ ($1$) in the subscript is to indicate that the pdf of the observations is assumed to be $f_0$ ($f_1$), so that the hypothesis in effect is the null hypothesis $H_0$ (the alternative hypothesis $H_1$, respectively). The two ARL metrics were introduced by Page~\cite{Page:B54} who formally defined them as $\ARL_i(w;h)\triangleq\EV_i[\mathcal{C}_{h}^w]$, $i=\{0,1\}$. It goes without saying that {\em both} ARLs are of interest, for either $\ARL_0(w;h)$ or $\ARL_1(w;h)$ alone does not tell the whole story as to the CUSUM Run Length's characteristics. However, given an initial score $w\in[0,h)$, a control limit $h>0$, and a particular observations model characterized by the densities $f_0(x)$ and $f_1(x)$, the evaluation of the ARLs is a major problem in applied sequential analysis, especially in quality control. To that end, a common practice in quality control has been to rely on the work of Page~\cite{Page:B54} who demonstrated that $\ARL_0(w;h)$ and $\ARL_1(w;h)$ satisfy certain integral (renewal) equations, which we state next.

For notational brevity, let $L_i(w;h)\triangleq\ARL_i(w;h)$, $i=\{0,1\}$. Then, according to Page~\cite{Page:B54}, we have
\begin{align}\label{eq:L-CS-eqn}
L_i(x;h)
&=
1+L_i(0;h)\,F_i(-x)+\int_{0}^{h} K_i(y-x)\,L_i(y;h)\,dy,\;i=\{0,1\},\; x\in[0,h),
\end{align}
where
\begin{align}\label{eq:Ki-def}
K_{i}(z)
&\triangleq
\frac{\partial}{\partial z}\Pr_{i}(\log\LR_{1}\le z),\; i=\{0,1\},\; z\in\mathbb{R},
\end{align}
i.e., $K_i(z)$ is the pdf of the log-likelihood ratio (log-LR) under the hypothesis $H_i$, $i=\{0,1\}$, and
\begin{align}\label{eq:Fi-def}
F_i(z)
&\triangleq
\int_{-\infty}^{z}K_{i}(x)\,dx,\; i=\{0,1\},\; z\in\mathbb{R},
\end{align}
i.e., $F_i(z)$ is the cumulative distribution function (cdf) of log-LR under the hypothesis $H_i$, $i=\{0,1\}$; in the quality control literature $K_i(z)$ is sometimes referred to as the ``frequency function'' (under the hypothesis $H_i$, $i=\{0,1\}$). Equations~\eqref{eq:L-CS-eqn} are renewal equations, and for either $i=\{0,1\}$ can be derived merely by conditioning on the first observation $X_1$.

Depending on the particular observations model given by the pair of pdf-s $f_i(x)$, $i=\{0,1\}$, equations~\eqref{eq:L-CS-eqn} may be recognized as either Fredholm (linear) integral equations of the second kind, or as Volterra integral equations, or as {\em delayed} Volterra integral equations. Regardless, a closed-form analytic solution is rarely an option. Hence the equations are usually solved {\em numerically}, and, as we mentioned in the introduction, the quality control literature is rife with numerical methods to solve specifically equations~\eqref{eq:L-CS-eqn}, assuming that the densities $f_i(x)$, $i=\{0,1\}$, the control limit $h>0$, and the headstart $w\in[0,h)$ are all given.

One of the first numerical methods to treat integral (renewal) equations akin to equations~\eqref{eq:L-CS-eqn} dates back to the work of Page~\cite{Page+Cox:MPCPhS1954}, and is based on a technique now known as the Markov Chain Monte Carlo (MCMC) method---a truly pioneering idea at that time. With regard to equations~\eqref{eq:L-CS-eqn} specifically, Page~\cite{Page:B54} also observed that the CUSUM chart is equivalent to repetitive application of the SPRT with the same initial score (headstart) and control boundaries at $0$ and at $h>0$. This equivalence is significant, for it provides a way to link together the performance of the CUSUM chart and that of the underlying repeated SPRT.

To be more specific, recall that the SPRT with control boundaries $a$ and $b$ ($a\le 0<b$, so that $a$ is the lower boundary and $b$ is the upper boundary) and initial score $w\in(a,b)$ is given by the stopping time
\begin{align*}
\mathcal{S}_{a,b}^w
&\triangleq
\min\{n\ge1\colon Z_n\not\in(a,b)\},
\end{align*}
where $Z_n\triangleq\sum_{i=1}^n\log\LR_i$ with $Z_0=w\in(a,b)$. Under either hypothesis $H_i$, $i=\{0,1\}$, the efficiency of the SPRT is customarily measured in terms of two functions: the ASN function and the OC function, defined, respectively, as $\ASN_i(w;a,b)\triangleq\EV_i[\mathcal{S}_{a,b}^w]$ and $\OC_i(w;a,b)=\Pr_i(Z_{\mathcal{S}_{a,b}^w}\le a)$, $i=\{0,1\}$. The decision made by the SPRT at termination is either ``accept $H_0$'' if the statistic $\{Z_n\}_{n\ge1}$ exits the interval $(a,b)$ through the lower boundary $a$, or ``reject $H_0$'' (i.e., ``accept $H_1$'') if the statistic $\{Z_n\}_{n\ge1}$ exits the interval $(a,b)$ through the upper boundary $b$. If the terminal decision is ``accept $H_0$'' (``reject $H_0$'') then the SPRT is referred to as an acceptance test (rejection test, respectively). We also note that, by definition, the OC function is the probability that the SPRT will terminate at the lower boundary $a$, under the appropriate hypothesis $H_i$, $i=\{0,1\}$.

The significance of the aforementioned equivalence between the CUSUM chart and a sequence SPRTs can now be made more clear: it allows to express the ARLs of the former through the ASNs and OCs of the latter. Specifically, for notational convenience, put $N_i(x;a,b)\triangleq\ASN_i(x;a,b)$ and $P_i(x;a,b)\triangleq\OC_i(w;a,b)$ for $i=\{0,1\}$. Since for either $i=\{0,1\}$ the SPRTs are applied {\em independently}, the number of acceptance tests before the first rejection test is a geometrically-distributed random variable. As a result, for each hypothesis $H_i$, $i=\{0,1\}$, the ARL of the CUSUM chart and the ASN and OC functions of the SPRT turn out to be connected through the relation
\begin{align}\label{eq:ARL-CS-ASN-OC-SPRT}
L_{i}(x;h)
&=
N_{i}(0;0,h)\,\frac{P_{i}(x;0,h)}{1-P_{i}(0;0,h)}+N_{i}(x;0,h),\;\;i=\{0,1\},
\end{align}
where $x\in[0,h)$ and $h>0$. A detailed derivation of the foregoing formula may be found, e.g., in~\cite[p.~387]{Tartakovsky+etal:Book2014}. It is now evident that the problem of computing the ARLs of the CUSUM chart boils down to the problem of computing the ASNs and the OCs of the underlying SPRT, and the latter problem, in turn, consists in recovering $N_0(x;a,b)$, $N_1(x;a,b)$, $P_0(x;a,b)$, and $P_1(x;a,b)$. With regard to computing the latter four quantities, Page~\cite{Page:B54} proved that they satisfy the following integral equations:
\begin{align}\label{eq:SPRT-NP-int-eqn}
\begin{split}
N_{0}(x;a,b)
&=
1+\int_{a}^{b} K_{0}(y-x)\, N_{0}(y;a,b)\,dy,\\
P_{0}(x;a,b)
&=
F_{0}(a-x)+\int_{a}^{b} K_{0}(y-x)\, P_{0}(y;a,b)\,dy,\\
N_{1}(x;a,b)
&=
1+\int_{a}^{b} K_{1}(y-x)\, N_{1}(y;a,b)\,dy,\\
P_{1}(x;a,b)
&=
F_{1}(a-x)+\int_{a}^{b} K_{1}(y-x)\, P_{1}(y;a,b)\,dy,
\end{split}
\end{align}
where $K_{i}(z)$ and $F_{i}(z)$ for $i=\{0,1\}$ are as in, respectively,~\eqref{eq:Ki-def} and~\eqref{eq:Fi-def} above. Just as equations~\eqref{eq:L-CS-eqn}, each of the foregoing four equations can also be derived by conditioning on the first observation $X_1$.

More importantly, just as equations~\eqref{eq:L-CS-eqn}, the four equations~\eqref{eq:SPRT-NP-int-eqn} are again integral (renewal) equations, so that, again, just as equations~\eqref{eq:L-CS-eqn}, they can rarely be solved analytically, compelling one to resort to the numerical solution. To explain the general idea behind any numerical method to solve equations~\eqref{eq:SPRT-NP-int-eqn}, consider the following generic integral equation
\begin{align}\label{eq:gen-int-eqn}
u(x)
&=
v(x)
+
\int_{a}^{b}K(y-x)\,u(y)\,dy,
\end{align}
where the unknown function is $u(x)$, and the nonhomogeneous term $v(x)$ as well as the integral equation's kernel $K(z)$ are given. The generic integral equation~\eqref{eq:gen-int-eqn} can be easily turned into any one of the four equations~\eqref{eq:SPRT-NP-int-eqn} merely by appropriately choosing $v(x)$ and $K(z)$. Indeed, setting $K(z)=K_{i}(z)$ with $K_{i}(z)$ given by~\eqref{eq:Ki-def} and $v(x)\equiv 1$ for all $x\in[a,b]$ gives the equation for $N_{i}(x;a,b)$, $i=\{0,1\}$. Likewise, keeping $K(z)=K_{i}(z)$ but instead setting $v(x)=F_i(a-x)$ with $F_{i}(z)$ as in~\eqref{eq:Fi-def} gives the equation for $P_{i}(x;a,b)$, $i=\{0,1\}$. Since equation~\eqref{eq:gen-int-eqn} combines all of the four equations~\eqref{eq:SPRT-NP-int-eqn}, any methodology to solve equation~\eqref{eq:gen-int-eqn} can be quickly adapted to any one of the four equations~\eqref{eq:SPRT-NP-int-eqn}. The main step of any (deterministic) numerical method to solve the generic integral equation~\eqref{eq:gen-int-eqn} is to linearize the integral in the right-hand side. This linearization can be performed, e.g., by means of a quadrature scheme, or using an interpolation method of some sort. The end-result of the linearization is that the original equation is reduced to a system of linear equations $\boldsymbol{u}=\boldsymbol{v}+\boldsymbol{K}\,\boldsymbol{u}$ which is then solved for $\boldsymbol{u}$ by standard linear-algebraic methods. Here $\boldsymbol{v}\triangleq[v(x_1),v(x_2),\ldots,v(x_n)]^\top$ where $\{x_j\}_{1\le j\le n}$ is a set of {\it a~priori} chosen $n\ge1$ discrete partition points of the interval $[a,b]$, i.e., $a\le x_1<x_2<\ldots<x_n\le b$. The $n\times n$ matrix $\boldsymbol{K}$ is a discrete equivalent of the integral operator
\begin{align}\label{eq:int-op}
\mathcal{K}\circ u
&\triangleq
\int_{a}^{b} K(y-x)\,u(y)\,dy,
\end{align}
and the elements of $\boldsymbol{K}$ are computed off the actual kernel $K(z)$ using the partition points $\{x_i\}_{1\le i\le n}$. If the approximation of $\mathcal{K}$ by $\boldsymbol{K}$ is sufficiently accurate, then the system $\boldsymbol{u}=\boldsymbol{v}+\boldsymbol{K}\,\boldsymbol{u}$ has a unique solution $\boldsymbol{u}\triangleq[u_1,u_2,\ldots,u_n]$, and it is reasonable to expect this solution to be close to the column-vector $[u(x_1),u(x_2),\ldots,u(x_n)]^{\top}$ of the actual values of the unknown function $u(x)$ at the partition points $\{x_j\}_{1\le j\le n}$. It is straightforward to see that $\boldsymbol{u}=(I-\boldsymbol{K})^{-1}\,\boldsymbol{v}$ where here and onward $I$ denotes the $n\times n$ identity matrix.

Going back to equations~\eqref{eq:SPRT-NP-int-eqn}, let $\boldsymbol{K}_{i}$, $i=\{0,1\}$, denote the matrix approximation of the integral operator~\eqref{eq:int-op} induced by the kernel $K_{i}(z)$, $i=\{0,1\}$, given by~\eqref{eq:Ki-def}. Suppose also that the corresponding partition points are $\{x_j\}_{1\le j\le n}$, and introduce $\boldsymbol{1}\triangleq[1,1,\ldots,1]^{\top}$ and $\boldsymbol{F}_{i}\triangleq[F_{i}(a-x_1),F_{i}(a-x_2),\ldots,F_{i}(a-x_n)]^{\top}$ for $i=\{0,1\}$. Then, by linearization, the four equations~\eqref{eq:SPRT-NP-int-eqn} are reduced to $\boldsymbol{N}_{0}=\boldsymbol{1}+\boldsymbol{K}_{0}\,\boldsymbol{N}_{0}$, $\boldsymbol{P}_{0}=\boldsymbol{F}_0+\boldsymbol{K}_{0}\,\boldsymbol{P}_{0}$,  $\boldsymbol{N}_{1}=\boldsymbol{1}+\boldsymbol{K}_{1}\,\boldsymbol{N}_{1}$, and $\boldsymbol{P}_{0}=\boldsymbol{F}_1+\boldsymbol{K}_{0}\,\boldsymbol{P}_{1}$. Here $\boldsymbol{N}_{i}$ and $\boldsymbol{P}_{i}$ are column-vectors comprised of approximate values of $N_{i}(x;a,b)$ and $P_{i}(x;a,b)$, respectively, evaluated at the partition nodes $\{x_j\}_{1\le j\le n}$. We are now in a position to make the following observation. While the system of linear equations for $\boldsymbol{N}_{0}$ and that for $\boldsymbol{P}_{0}$ have different nonhomogeneous terms, they both have the same matrix of coefficients $I-\boldsymbol{K}_0$. As a result, both systems can be (and should be) solved simultaneously by combining the nonhomogeneous terms into the $n\times 2$ matrix $[\boldsymbol{1},\boldsymbol{F}_0]$ and solving the system $(I-\boldsymbol{K}_0)\,[\boldsymbol{N}_0,\boldsymbol{P}_0]=[\boldsymbol{1},\boldsymbol{F}_0]$. Likewise, the system for $\boldsymbol{N}_{1}$ and that for $\boldsymbol{P}_{1}$ can be solved in exactly the same manner. There is no question that grouping the right-hand sides of any two or more different systems of linear equations with the same matrix of coefficients allows to cut down the overall number of operations needed to solve all of the systems. This a basic fact taught in any course on elementary linear algebra. The problem, however, is that of the four equations~\eqref{eq:SPRT-NP-int-eqn}, the top two (which correspond to the null hypothesis) and the bottom two (which correspond to the alternative hypothesis) appear to be unrelated, and therefore have to be solved {\em separately}. The main result of this paper is to prove otherwise. Specifically, it turns out that $K_1(z)$ and $K_0(z)$ {\em are} connected, and the connection is simple and allows one to show that the four equations~\eqref{eq:SPRT-NP-int-eqn} are all instances of the same single equation involving the same kernel and parameterized only by the nonhomogeneous term. As a result, all four equations~\eqref{eq:SPRT-NP-int-eqn} can be solved {\em simultaneously} by grouping the nonhomogeneous terms together, just as we described above for the generic equation~\eqref{eq:gen-int-eqn}. This would clearly lead to a reduction of the computational burden required to approximately recover $N_{i}(x;a,b)$ and $P_{i}(x;a,b)$ as $\boldsymbol{N}_{i}$ and $\boldsymbol{P}_{i}$, respectively. The specifics are discussed in the next section.

\section{The Main Result and Its Discussion}
\label{sec:main-results}

We begin with an observation that will be key to obtain a link between $K_0(z)$ and $K_1(z)$ given by~\eqref{eq:Ki-def}, and subsequently establish the main result of this paper. Let $P_i^{\LR}(t)\triangleq\Pr_i(\LR_1\le t)$, $t\ge0$, $i=\{0,1\}$, denote the cdf of the LR under the hypothesis $H_i$, $i=\{0,1\}$, respectively. Since the LR is the Radon--Nikod\'{y}m derivative of the probability measure $\Pr_1$ with respect to the probability measure $\Pr_0$ (the two measures are assumed to be mutually absolutely continuous), one can deduce the following result.
\begin{lemma}\label{lem:change-of-measure-id1}
$dP_1^{\LR}(t)=t\,dP_0^{\LR}(t)$, $t\ge0$.
\end{lemma}

This result is nothing but Wald's~\cite{Wald:Book47} likelihood ratio identity (see also, e.g.,~\cite[p.~13]{Siegmund:Book85},~\cite[p.~4]{Woodroofe:Book82},~\cite{Lai:SA2004}, or~\cite[Theorem~2.3.3,~p.~32]{Tartakovsky+etal:Book2014}), and can be obtained from the following argument:
\begin{align*}
\begin{split}
dP_{1}^{\LR}(t)
&\triangleq d\Pr_{1}(\LR_{1}\le t)\\
&=
d\Pr_{1}(X_1\le \LR_{1}^{-1}(t))\\
&=
\LR_{1}(\LR_{1}^{-1}(t))\,d\Pr_0(X_{1}\le \LR_{1}^{-1}(t))\\
&=
t\,d\Pr_{0}(\LR_{1}\le t)\\
&=
t\,dP_{0}^{\LR}(t),
\end{split}
\end{align*}
whence $dP_1^{\LR}(t)=t\,dP_0^{\LR}(t)$, $t\ge0$, as needed; cf.~\cite{Polunchenko+etal:SA2014,Polunchenko+etal:ASMBI2014} and~\cite[Chapter~3]{Du:PhD-Thesis2015}.

As an immediate implication of Lemma~\ref{lem:change-of-measure-id1}, observe that since
\begin{align*}
\begin{split}
K_{i}(z)
&\triangleq
\frac{d}{dz}\Pr_{i}(\log\LR_1 \le z)\\
&=
\frac{d}{dz}\Pr_{i}(\LR_1\le e^z)\\
&=
\frac{d}{dz}P_{i}^{\LR}(e^z),\; i=\{0,1\},
\end{split}
\end{align*}
it follows that $K_{1}(z)=e^{z}\,K_{0}(z)$, $z\in\mathbb{R}$. Now, setting $z=y-x$, the following can be seen to hold true.
\begin{lemma}\label{lem:change-of-measure-id2}
$e^{-y}\,K_{1}(y-x)=e^{-x}\,K_{0}(y-x)$, $x,y\in\mathbb{R}$.
\end{lemma}

The foregoing lemma is the main result of this paper. It is an obvious extension of the results obtained previously in~\cite{Polunchenko+etal:SA2014,Polunchenko+etal:ASMBI2014} for the Generalized Shiryaev--Roberts procedure. As simple as it may seem, the established connection between $K_1(z)$ and $K_0(z)$ has far-reaching consequences. We shall now elaborate on this at greater length.

At the very least Lemma~\ref{lem:change-of-measure-id2} provides a ``shortcut'' to derive a formula for ${K}_1(z)$ from that for ${K}_{0}(z)$, or the other way around---whichever one of the two is found first. To illustrate this point, suppose that
\begin{align*}
f_0(x)
&=
\frac{1}{\sqrt{2\,\pi}}\,e^{-\tfrac{x^2}{2}}
\;\text{and}\;
f_1(x)
=
\frac{1}{\sqrt{2\,\pi}}\,e^{-\tfrac{(x-\theta)^2}{2}},
\end{align*}
where $x\in\mathbb{R}$ and $\theta\neq 0$, a known parameter (which is the ``off-target'' mean level). This basic Gaussian model is the standard ``testbed'' model widely used in the literature for demonstrational purposes. Under this model it is direct to see that the log-LR is of the form
\begin{align}\label{eq:LR-Gauss}
\log\LR_n
&\triangleq
\log\frac{f_1(X_n)}{f_0(X_n)}=
\theta\,X_n-\frac{\theta^2}{2},
\end{align}
whence
\begin{align*}
K_0(z)
&=
\frac{1}{\sqrt{2\,\pi\,\theta^2}}\,\exp\left\{-\frac{1}{2\theta^2}\left(z+\frac{\theta^2}{2}\right)^2\right\},\; z\in\mathbb{R},
\end{align*}
i.e., $K_0(z)$ is the pdf of a Gaussian distribution with mean $-\theta^2/2\,(<0)$ and variance $\theta^2\,(\neq 0)$. As a result, from $K_1(z)=e^{z}\,K_0(z)$ we obtain that
\begin{align*}
K_1(z)
&=
\frac{1}{\sqrt{2\,\pi\,\theta^2}}\,\exp\left\{-\frac{1}{2\theta^2}\left(z-\frac{\theta^2}{2}\right)^2\right\},\; z\in\mathbb{R},
\end{align*}
i.e., $K_1(z)$ is the pdf of a Gaussian distribution with mean $\theta^2/2\,(>0)$ and variance $\theta^2\,(\neq 0)$. This is exactly what $K_1(z)$ is supposed to be for the Gaussian model at hand. We stress that the formula for $K_1(z)$ was {\em not} obtained from~\eqref{eq:LR-Gauss}, i.e., from the log-LR formula: we used the log-LR formula~\eqref{eq:LR-Gauss} to recover $K_0(z)$ only, and then with $K_0(z)$ expressed explicitly, we exploited the identity $K_1(z)=e^{z}\,K_0(z)$ to find $K_1(z)$. When $f_i(x)$, $i=\{0,1\}$, are complicated, and the log-LR is not a simple function, obtaining both $K_i(z)$, $i=\{0,1\}$, in a closed-form {\em directly} from the log-LR formula may be rather ``calculusy''. It is in such cases that appealing instead to the identity $K_1(z)=e^{z}\,K_0(z)$ may prove especially advantageous, for the calculus involved is effectively half that required to get $K_1(z)$ directly from the log-LR formula.

More importantly, observe that in view of Lemma~\ref{lem:change-of-measure-id2}, the aforementioned four integral equations~\eqref{eq:SPRT-NP-int-eqn} on $N_i(x;a,b)$ and $P_i(x;a,b)$, $i=\{0,1\}$, can be rewritten as follows:
\begin{align*}
\begin{split}
N_0(x;a,b)
&=
1+\int_{a}^b K_0(y-x)\, N_0(y;a,b)\,dy,\\
e^x\,N_1(x;a,b)
&=
e^x+\int_{a}^b K_0(y-x)\,[e^{y}\,N_1(y;a,b)]\,dy,\\
P_0(x)
&=
F_0(a-x)+\int_{a}^b K_0(y-x)\,P_0(y;a,b)\,dy,\\
e^{x}\,P_1(x)
&=
e^{x}\,F_1(a-x)+\int_a^b K_0(y-x)\,[e^{y}\,P_1(y;a,b)]\,dy,
\end{split}
\end{align*}
so that the kernel $K_{1}(z)$ is eliminated entirely, and all of the equations turn out to involve only the kernel $K_0(z)$. It is now evident that the ASN and OC functions of the SPRT under the null and under the alternative are {\em all} governed by the same {\em one} integral equation but with different nonhomogeneous terms: the equation for $N_0(x;a,b)$ has 1 as its nonhomogeneous term, the equation for $P_0(x;a,b)$ has $F_0(a-x)$ as its nonhomogeneous term, the equation for $e^x\,N_1(x;a,b)$ has $e^x$ as its nonhomogeneous term, and the equation for $e^x\,P_1(x;a,b)$ has $e^{x}\,F_1(a-x)$ as its nonhomogeneous term. We note also that the latter two equations are to be solved not for $N_1(x;a,b)$ and $P_1(x;a,b)$, but for $e^x\, N_1(x;a,b)$ and $e^x\, P_1(x;a,b)$, respectively, and the exponential factor present in the obtained solutions is taken care of once $e^x\, N_1(x;a,b)$ and $e^x\, P_1(x;a,b)$ are found. Therefore, Lemma~\ref{lem:change-of-measure-id2} allows to find the ASNs and OCs of the SRPT under the null and under the alternative {\em simultaneously} in the manner that was explained at the end of Section~\ref{sec:preliminaries}, i.e., by simply grouping the nonhomogeneous terms into one matrix. Furthermore, via the relation~\eqref{eq:ARL-CS-ASN-OC-SPRT} the in- and out-of-control ARLs of the CUSUM chart can also be computed {\em concurrently}, thereby making a more efficient use of the computational resources available. This is the primary {\em practical} benefit of Lemma~\ref{lem:change-of-measure-id2}, i.e., the main result of this paper.

To provide yet another illustration of the practical benefits of Lemma~\ref{lem:change-of-measure-id2}, consider the problem of computing the entire distribution of the CUSUM Run Length under the hypothesis $H_i$, $i=\{0,1\}$. The respective integral equations and recurrences were obtained, e.g., by Ewan and Kemp~\cite{Ewan+Kemp:B1960} and then also by Woodall~\cite{Woodall:T1983}. See also, e.g., Waldmann~\cite{Waldmann:T1986}. Specifically, capitalizing on the fact that the CUSUM chart is a sequence of SPRTs, for each hypothesis $H_i$, $i=\{0,1\}$, Woodall~\cite{Woodall:T1983} expressed the distribution of the CUSUM Run Length through that of the Run Length of the SPRT. See~\cite[p.~296]{Woodall:T1983}. The distribution of the SPRT, in turn, is found using repetitive application of the linear integral operator
\begin{align}\label{eq:int-op-i}
\mathcal{K}_{i}\circ u
&\triangleq
\int_{a}^{b} K_{i}(y-x)\,u(y)\,dy,
\end{align}
where $i=\{0,1\}$. Therefore, from Lemma~\ref{lem:change-of-measure-id2} it is easy to see that the distribution of the SPRT Run Length under the hypothesis $H_0$ can be found {\em concurrently} with the distribution under the hypothesis $H_1$.

We would like to conclude this section with two remarks. First, note that the integral operator~\eqref{eq:int-op-i} was involved in all of the integral equations we considered. As a matter of fact, the integral operator~\eqref{eq:int-op-i} can be used as a universal index that summarizes the overall performance of the chart, whatever the specific metric be. By way of example, if one were interested in computing the higher-order $H_i$-moments of the CUSUM Run Length, i.e., the quantities $\mu_{i}^{(k)}\triangleq\EV_{i}[(\mathcal{C}_h^w)^k]$, $i=\{0,1\}$, $k=0,1,\ldots$, then the equivalence of the CUSUM chart to sequential application of the SPRT would again allow to express $\mu_{i}^{(k)}$ through the corresponding moments of the SPRT, and the latter moments would again satisfy integral equations similar to equations~\eqref{eq:SPRT-NP-int-eqn}. See, e.g.,~\cite{Ewan+Kemp:B1960}. Therefore, Lemma~\ref{lem:change-of-measure-id2} would again allow to lessen the computational cost, for it would allow to compute $\mu_{1}^{(k)}$ and $\mu_{0}^{(k)}$ {\em concurrently} for any fixed $k=0,1,\ldots$. Second, note that for Page's~\cite{Page:B54} symmetric two-sided CUSUM scheme, Lemma~\ref{lem:change-of-measure-id2} would allow for an even greater reduction of the computational complexity.

\section{Conclusion}
\label{sec:conclusion}

As part of the author's ongoing effort to bridge the gap between the theory and application of sequential analysis, this work sought to build on to the results obtained previously in~\cite{Polunchenko+etal:SA2014,Polunchenko+etal:ASMBI2014} and in~\cite[Chapter~3]{Du:PhD-Thesis2015}, and provide an example of an alternative, more {\em applicative} use of one of the central techniques of {\em theoretical} sequential analysis---Wald's~\cite{Wald:Book47} likelihood ratio identity. Specifically, by virtue of the latter we obtained a connection between a broad range of {\em in-control} characteristics of the CUSUM chart and their {\em out-of-control} counterparts. The obtained connection relates directly the integral equations on the {\em in-control} characteristics and the integral equations on the corresponding {\em out-of-control} characteristics. On a practical level, this relationship allows the {\em in-} and {\em out-of-control} characteristics to be recovered {\em simultaneously} as solutions of the respective integral equations, thereby improving the efficiency of any numerical method one may employ to compute the performance of the CUSUM chart as well as that of the SPRT.

\section*{Acknowledgements}
The author is thankful to the Editor-in-Chief, Dr. Fabrizio Ruggeri of the Institute of Applied Mathematics and Information Technology, Italian National Research Council (CNR IMATI), and to the three anonymous referees whose constructive feedback helped improve the quality of the manuscript and shape its current form.

The author's effort was partially supported by the Simons Foundation via a Collaboration Grant in Mathematics under Award \#\,304574.

\singlespacing

\end{document}